\documentclass[11pt,twoside]{article} % Specifies the document style.
\usepackage{times,fancyhdr}
\usepackage{cite}
\usepackage{graphicx}

\usepackage{hhline}

\def\beq{\begin{equation}}
\def\eeq{\end{equation}}

\def\s{{\,\rm s}}
\def\g{{\,\rm g}}
\def\K{\,{\rm K}}
\def\eV{\,{\rm eV}}
\def\keV{\,{\rm keV}}
\def\MeV{\,{\rm MeV}}
\def\GeV{\,{\rm GeV}}

\def\s{\,{\rm s}}
\def\cm{\,{\rm cm}}

\def\bn{\,{\rm barn}}

\def\to{\rightarrow}

\newcommand{\lsim}{\lower .7ex\hbox{$\;\stackrel{\textstyle <}{\sim}\;$}}
\newcommand{\gsim}{\lower .7ex\hbox{$\;\stackrel{\textstyle >}{\sim}\;$}}
\def\sv{\left<\sigma v\right>}
\newcommand{\fr}[3]{\left(\frac{#1}{#2}\right)^{\!\! #3}}
\def\anti#1#2#3{(\bar{#1}\bar{#2}\bar{#3})}
\def\OHe{$O$-$Helium$ }

\date{}

\title{Stable quarks of the 4th family?} % Declares the document's title.
\author{K. Belotsky$^1$, M. Khlopov$^{1,2}$, K. Shibaev$^1$\\
\\
{\small 1) Moscow Engineering Physics Institute, Moscow, Russia;}\\
{\small Center for Cosmoparticle physics ``Cosmion'', Moscow, Russia}\\
{\small 2) APC laboratory,
Paris, France} }

\begin{document} % End of preamble and beginning of text.

%\pagestyle{fancy}
%\fancyhead{} % clear all header fields
%\fancyhead[EC]{K. Belotsky, M. Khlopov, K. Shibaev}
%\fancyhead[EL,OR]{\thepage}
%\fancyhead[OC]{Stable quarks of the 4th family?}
%\fancyfoot{} % clear all footer fields
%\renewcommand\headrulewidth{0.5pt}
%\addtolength{\headheight}{2pt} % make space for the rule

\maketitle % Produces the title.

\begin{abstract}
Existence of metastable quarks of new generation can be embedded
into phenomenology of heterotic string together with new long range
interaction, which only this new generation possesses. We discuss
primordial quark production in the early Universe, their successive
cosmological evolution and astrophysical effects, as well as
possible production in present or future accelerators. In case of a
charge symmetry of 4th generation quarks in Universe, they can be
stored in neutral mesons, doubly positively charged baryons, while
all the doubly negatively charged "baryons" are combined with He-4
into neutral nucleus-size atom-like states. The existence of all
these anomalous stable particles may escape present experimental
limits, being close to present and future experimental test. Due to
the nuclear binding with He-4 primordial lightest baryons of the 4th
generation with charge $+1$ can also escape the experimental upper
limits on anomalous isotopes of hydrogen,
being compatible with upper limits on anomalous lithium.
While 4th quark hadrons
are rare, their presence may be nearly detectable in cosmic rays,
muon and neutrino fluxes and cosmic electromagnetic spectra. In case
of charge asymmetry, a nontrivial solution for the problem of dark
matter (DM) can be provided by excessive (meta)stable anti-up quarks
of 4th generation, bound with He-4 in specific nuclear-interacting
form of dark matter. Such candidate to DM is surprisingly close to
Warm Dark Matter by its role in large scale structure formation.
It catalyzes primordial heavy element production in Big Bang
Nucleosynthesis and new types of nuclear transformations around us.
\end{abstract}

\section{\label{introduction}Introduction}

The question about existence of new quarks and/or leptons is among
the most important in the modern particle physics.
Possibility of existence of new (meta)stable quarks which form
new (meta)stable hadrons is of special interest.
New stable
hadrons can play the role of strongly interacting dark matter
\cite{Dimopoulos:1989hk,Starkman,McGuire:2001qj}. This question is
believed to find solution in the framework of future Grand Unified
Theory. A strong motivation for existence of new long-living hadrons
comes from a possible solution \cite{Dvali} of the "doublet-triplet
splitting" problem in supersymmetric GUT models. Phenomenology of
string theory offers another motivation for new long lived hadrons.

A natural extension of the Standard model can lead in the heterotic
string phenomenology to the prediction of fourth generation of
quarks and leptons \cite{Shibaev,Sakhenhance} with a stable 4th
neutrino \cite{Fargion99,Grossi,Belotsky,BKS}. The comparison
between the rank of the unifying group $E_{6}$ ($r=6$) and the rank
of the Standard model ($r=4$) can imply the existence of new
conserved charges. These charges can be related with (possibly
strict) gauge symmetries. New strict gauge U(1) symmetry (similar to
U(1) symmetry of electrodynamics) is excluded for known particles
but is possible, being ascribed to the fermions of 4th generation
only. This provides theoretic motivation for a stability of the
lightest fermion of 4th generation, assumed to be neutrino. Under
the condition of existence of strictly conserved charge, associated
to 4th generation, the lightest 4th generation quark $Q$ (either $U$
or $D$) can decay only to 4th generation leptons owing to GUT-type
interactions, what makes it sufficiently long living.

Whatever physical reason was for a stability of new hypothetical particles,
it extends potential for testing respective hypothesis due to its implications
in cosmology.
Especially rich in this sense is a hypothesis on (meta)stable quarks of new family.
It defines the goal of current work.

As we will show, in the case when 4th generation possesses strictly
conserved $U(1)$-gauge charge (which will be called $y$-charge), 4th
generation fermions are the source of new interaction of Coulomb
type (which we'll call further $y$-interaction). It can be crucial
for viability of model with equal amounts of 4th generation quarks
and antiquarks in Universe. The case of cosmological excess of 4th
generation antiquarks offers new form of dark matter with a very
unusual properties. Owing to strict conservation of $y$-charge, this
excess should be compensated by excess of 4th generation neutrinos.

Recent analysis \cite{Okun} of precision data on the Standard model
parameters admits existence of the 4th generation particles, satisfying direct experimental constraints
which put lower limit 220 GeV for the mass of lightest quark \cite{PDG}.

If the lifetime of the lightest 4th generation quark exceeds the age of the Universe,
primordial $Q$-quark (and $\bar Q$-quark) hadrons should be present in the modern matter.
If this lifetime is less than the age of the Universe, there should be no
primordial 4th generation quarks, but they can be produced
in cosmic ray interactions and be present in cosmic ray fluxes.
The search for this quark is a challenge for the present and future accelerators.

In the present work we will assume that up-quark of 4th generation ($U$) is lighter
than its down-quark ($D$). The opposite assumption is found to be virtually excluded, if $D$ is stable.
The reason is that $D$-quarks might form stable hadrons with electric charges $\pm 1$
($(DDD)^-$, $(\bar Du)^+$, $\anti DDD^+$), which eventually form hydrogen-like atoms
(hadron $(DDD)^-$ is combined with $^4He^{++}$ into $+1$ bound state),
being strongly constrained in surrounding matter.
It will become more clear from consideration of $U$-quark case, presented below.

The following hadron states containing (meta)stable $U$-quarks (U-hadrons) are expected to be (meta)stable
and created in early Universe:
``baryons'' $(Uud)^+$, $(UUu)^{++}$, $(UUU)^{++}$;
``antibaryons'' $\anti UUU^{--}$, $\anti UUu^{--}$, meson $(\bar U u)^0$.
The absence in the Universe of the states $\anti Uud$, $(U\bar u)$ containing light antiquarks are suppressed because of baryon asymmetry.
Stability of double and triple $U$ bound states $(UUu)$, $(UUU)$ and $\anti UUu$, $\anti UUU$ is provided
by the large chromo-Coulomb binding energy ($\propto \alpha_{QCD}^2 \cdot m_Q$)
\cite{Glashow:2005jy,Fargion:2005xz}.
Formation of these states in
particle interactions at accelerators and in cosmic rays is
strongly suppressed, but they can form in early Universe and
cosmological analysis of their relics can be of great importance
for the search for 4th generation quarks.

We analyze the mechanisms of production of metastable $U$ (and
$\bar U$) hadrons in the early Universe, cosmic rays and
accelerators and point the possible signatures of their
existence.

We'll show that in case of charge symmetry of U-quarks in Universe,
a few conditions play a crucial role for viability of the model. An
electromagnetic binding of $\anti UUU^{--}$ with $^4He^{++}$ into
neutral nucleus-size atom-like state (\OHe) should be accompanied by
a nuclear fusion of $(Uud)^+$ and $^4He^{++}$ into lithium-like
isotope $[^4He(Uud)]$ in early Universe.
The realization of such %$(Uud)^+$ and $^4He^{++}$
a fusion requires
a marginal supposition concerning respective cross section.
Furthermore, assumption of $U(1)$-gauge nature of the charge,
associated to U-quarks, is needed to
avoid a problem of overproduction of anomalous isotopes by means of
an $y$-annihilation of U-relics ($[^4He(Uud)]$, $(UUu)$, $(UUU)$, $^4He\anti UUU$, $^4He\anti UUu$, $(\bar U u)$).
Residual amount of U-hadrons with respect to baryons in this case
is estimated to be less than $10^{-10}$ in Universe in toto and less than $10^{-20}$ at the Earth.

A negative sign charge asymmetry of U-quarks in Universe
can provide a nontrivial solution for dark matter (DM) problem.
For strictly conserved charge such asymmetry in $\bar U$ implies
corresponding asymmetry in leptons of 4th generation.
In this case the most of $\bar U$ in Universe are contained in \OHe states
$[^4He\anti UUU]$ and minor part of them in mesons $\bar Uu$.
On the other hand the set of
direct and indirect effects of relic U-hadrons existence provides the test
in cosmic ray and underground experiments which can be decisive
for this hypothesis.
The main observational effects for asymmetric case
do not depend on the existence of $y$-interaction.

The structure of this paper is as the following. Section 2 is
devoted to the charge symmetric case of U-quarks. Cosmological
evolution of U-quarks in early Universe is considered in subsection
2.1, while in subsection 2.2 the evolution and all possible effects
of U-quarks existence in our Galaxy are discussed. The case of
charge asymmetry of quarks of 4th generation in Universe is
considered in Section 3. Section 4 is devoted to the questions of
the search for U-quarks at accelerators. We summarize the results of
our present study, developing earlier investigations \cite{BFK,KPS},
 in Conclusion.

%%%%%%%%%%%%%%%%%%%%%%%%%%%%%%%%%%%%%%%%%%%%%%%%%%%%%%%%%%%%%%%%%%%%%%%%%%%%%%%%%%%
%%%%%%%%%%%%%%%%%%%%%%%%%%%%%%%%%%%%%%%%%%%%%%%%%%%%%%%%%%%%%%%%%%%%%%%%%%%%%%%%%%%
\section{Charge symmetric case of U-quarks}

\subsection{\label{primordial} Primordial $U$-hadrons from Big Bang Universe}

\textbf{\indent Freezing out of U-quarks}\\*
In the early Universe at temperatures highly above their masses
fermions of 4th generation were in thermodynamical equilibrium
with relativistic plasma.
When in the course of expansion
the temperature $T$ falls down below the mass of the lightest $U$-quark, $m$,
equilibrium concentration of quark-antiquark pairs of 4th generation
is given by
\beq
n_4 = g_4 \fr{Tm}{2\pi}{3/2} \exp{(-m/T)},
\label{equil}
\eeq
where $g_4=6$ is the effective number of their spin and colour degrees of freedom. We use the units
$\hbar = c = k =1$ throughout this paper.

The expansion rate of the Universe at RD-stage is given by the expression
\beq
H=\frac{1}{2t}=\sqrt{ \frac{4\pi^3g_{tot}}{45}}\frac{T^2}{m_{Pl}}\approx 1.66\,g_{tot}^{1/2}\frac{T^2}{m_{Pl}},
\label{Hubble}
\eeq
where temperature dependence follows from the expression for critical density of the Universe
$$\rho_{crit}=\frac{3H^2}{8\pi G}=g_{tot}\frac{\pi^2}{30}T^4.$$
When it
starts to exceed the rate of quark-antiquark annihilation
\beq
R_{ann}=n_4 \sv,
\label{annih}
\eeq
in the period, corresponding to $T=T_f<m$, quarks of 4th generation freeze out,
so that their concentration does not follow the equilibrium distribution Eq.(\ref{equil}) at $T<T_f$.
For a convenience we introduce the variable
\beq
r_4=\frac{n_4}{s},
\label{r4}
\eeq
where
\beq
s=\frac{2\pi^2g_{tot\,s}}{45}T^3\approx 1.80\,g_{tot\,s}n_{\gamma}\approx 1.80\,g_{tot\,s}^{mod}\eta^{-1}n_B
\label{entropy}
\eeq
is the entropy density of all matter.
In Eq.(\ref{entropy}) $s$ was expressed through the thermal photon number density
$n_{\gamma}=\frac{2\zeta(3)}{\pi^2}T^3$ and also through the baryon number density $n_B$, for which at the modern
epoch we have $n_B^{mod}/n_{\gamma}^{mod}\equiv \eta\approx 6\cdot 10^{-10}$.

Under the condition of entropy conservation in the Universe, the number density
of the frozen out particles can be simply found
for any epoch through the corresponding thermal photon number density $n_{\gamma}$. Factors $g_{tot}$
and $g_{tot\,s}$ take into account the contribution of all particle species and are defined as
$$g_{tot}=\sum_{i=bosons} g_i \fr{T_i}{T}{4}
+\frac{7}{8}\sum_{i=fermions} g_i \fr{T_i}{T}{4}$$
and
$$g_{tot\,s}=\sum_{i=bosons} g_i \fr{T_i}{T}{3}
+\frac{7}{8}\sum_{i=fermions} g_i \fr{T_i}{T}{3},$$
where $g_i$ and $T_i$ are the number of spin degrees of freedom and temperature of ultrarelativistic bosons
or fermions. For epoch $T\ll m_e\approx 0.5\MeV$ it is assumed that only
photons and neutrinos with $T_{\nu}=(4/11)^{1/3}T$ give perceptible contribution into energy (until the end of RD-stage)
and entropy (until now) densities so one has
\beq
g_{tot\,s}^{mod}\approx 3.91 \qquad g_{tot}^{mod}\approx 3.36.
\label{gmod}
\eeq
For modern entropy density we have $s_{mod}\approx 2890 \cm^{-3}$.

From the equality of the expressions Eq.(\ref{Hubble}) and Eq.(\ref{annih}) one gets
$$m/T_f\approx42+\ln(g_{tot}^{-1/2}m_pm\sv)$$
with $m_p$ being the proton mass and obtains, taking
$\sv \sim \frac{\alpha^2_{QCD}}{m^2}$ and $g_{tot}(T_f)=g_{tot\,s}(T_f)= g_f\approx80-90$,
$$T_f \approx m/30$$
and
\beq
r_4=\frac{H_f}{s_f\sv}\approx \frac{4}{g_f^{1/2}m_{Pl}T_f\sv} \approx 2.5 \cdot 10^{-14}\frac{m}{250
{\GeV}}.
\label{freez2}
\eeq
Index "f" means everywhere that the corresponding quantity is taken at $T=T_f$.
Note, that the result Eq.(\ref{freez2}), obtained in approximation
of "instantaneous" freezing out, coincides with more accurate one if $\sv$ and
$g_f$ can be considered (as in given case) to be constant.
Also it is worth to emphasize,
that given estimation for $r_4$ relates
to only 4th quark or 4th antiquark abundances, assumed in this part to be equal to each other.

Note that if $T_f > \Delta = m_D - m$, where $m_D$ is the mass of $D$-quark
(assumed to be heavier, than $U$-quark) the frozen out concentration of 4th generation quarks
represent at $T_f > T> \Delta$ a mixture of nearly equal amounts of
$U \bar U$ and $D \bar D$ pairs.

At $T < \Delta$ the equilibrium ratio
$$\frac{D}{U} \propto \exp{\left(-\frac{\Delta}{T}\right)}$$
is supported by weak interaction, provided that
$\beta$-transitions $(U \rightarrow D)$ and $(D \rightarrow U)$ are
in equilibrium.
The lifetime of $D$-quarks, $\tau$, is also determined by
the rate of weak $(D \rightarrow U)$ transition, and at $t \gg \tau$
all the frozen out $D \bar D$ pairs should decay to $U \bar U$ pairs.

%%%%%%%%%%%%%%%%%%%%%%%%%%%%%%%%%%%%%%%%%%%

At the temperature $T_f$ annihilation of U-quarks to gluons and to pairs of light quarks
$U \bar U \rightarrow gg ,q\bar q$ terminates
and $U \bar U$ pairs are frozen out. The frozen out
concentration is given by Eq.(\ref{freez2}).
Even this value of primordial concentration
of $U$-quarks with the mass $m = 250$ GeV would lead to the
contribution into the modern density $2mr_4s_{mod}$, which is by an order of
magnitude less than the baryonic density, so that
in the charge symmetric case  $U$-quarks can
not play a significant dynamical role in the modern Universe.

The actual value of primordial $U$-particle concentration should be
much smaller due to QCD, hadronic and radiative recombination, which
reduce the abundance of frozen out $U$-particles. $y$-Interaction
can play essential role in successive evolution to be considered. It
accounts for radiative recombination and plays crucial role in
galactic evolution of U-hadrons. So, it will be included into
further consideration which will be carried out for both sub-cases
(with and without $y$-interaction).

%\newline
\textbf{QCD recombination}\\*
At $$T \le I_1 = m \bar \alpha^2/4 = 3.2 {\GeV}
\frac{m}{250{\GeV}},$$ where $\bar \alpha = 0.23$ accounts for
joint effect of Coulomb-like attraction due to QCD and $y$-interactions,
formation of bound $(U \bar U)$
states is possible, in which frozen out Heavy quarks and
antiquarks can annihilate. Effect of $y$-interaction is not essential here.

Note that at $T \le I_1$ rate of $(U \bar
U)$ annihilation in bound systems exceeds the rate of "ionization"
of these systems by quark gluon plasma. So the rate of QCD
recombination, given by \cite{Fargion:2005xz,BFK}
\beq \sv \approx
\left(\frac{16\pi}{3^{5/2}}\right) \cdot \frac{\bar
\alpha}{T^{1/2}\cdot m_U^{3/2}},
\label{svUbar}
\eeq
is the rate, with which abundance of frozen out $U$-quarks decreases.

The decrease of $U$-hadron abundance owing to $U \bar U$
recombination is governed by the equation
\beq
\frac{dn_4}{dt} = -3Hn_4-n_4^2 \cdot \sv.
\label{hadrecomb1}
\eeq
Using notation Eq.(\ref{r4}) and relation
\beq
-dt=\frac{dT}{HT},
\eeq
which follows from Eq.(\ref{Hubble}) and is true as long as $g_{tot}\approx const$,
Eq.(\ref{hadrecomb1}) is reduced to
\beq
dr_4=r_4^2\cdot s HT\sv dT,
\label{hadrecomb}
\eeq
where $sHT=\sqrt{\pi g/45}$ with $g\equiv g_{tot\,s}^2/g_{tot} =({\rm for}\,\, T>m_e)=g_{tot\,s}=g_{tot}$.

At $T_0 = I_1 > T > T_{QCD}=T_1$, assuming in this period
$g=const=g_f \approx 17$, the solution of Eq.(\ref{hadrecomb}) is
given by \beq r_4 = \frac{r_0}{1 + r_0 \sqrt{\frac{\pi g_f}{45}}\,
m_{Pl} \int^{T_0}_{T_1}\sv dT} \approx 0.16\,
\left(\frac{m}{I_1}\right)^{1/2} \frac{m}{\bar \alpha m_{Pl}}\approx
\label{radrecsolq} \eeq
$$\approx 1.6 \cdot 10^{-16} \frac{m}{250\GeV}.$$
It turns to be independent on the
frozen out concentration $r_0$ given by Eq.(\ref{freez2}).

At $T<I_{UU} \le m \tilde \alpha^2/4 = 1.6 \GeV
\frac{m}{250\GeV},$ where effective constant $\tilde
\alpha=C_F\alpha_s - \alpha_y \sim (4/3) \cdot 0.144 - 1/30 =
0.16$ accounts for repulsion of the same sign $y$-charges,
reactions $U + U \rightarrow (UU) + g$ and $U + (UU) \rightarrow
(UUU) + g$ can lead to formation $(UU)$-diquark and colorless
$(UUU)$ "hadron" (as well as similar $\bar U$ bound states) in
quark gluon plasma \cite{Glashow:2005jy,Fargion:2005xz}. However,
disruption of these systems by gluons in inverse reactions
prevents their effective formation at $T \gsim I_{UU}/30$
\cite{Fargion:2005xz}. Therefore, such systems of $U$ quarks with
mass $m < 700 \GeV$ are not formed before QCD phase transition.

%\newline
\textbf{Hadronic recombination}\\*
After QCD phase transition at $T = T_{QCD} \approx 150$MeV quarks
of 4th generation combine with light quarks into $U$-hadrons. In
baryon asymmetrical Universe only excessive valence quarks should
enter such hadrons.
Multiple U states formation can start only in processes of hadronic recombination
for U-quark mass $m<700$ GeV what is discussed below.

As it was revealed in \cite{Shibaev,Sakhenhance}
in the collisions of such mesons and baryons recombination
of $U$ and $\bar U$ into unstable $(U \bar U)$ "charmonium -like" state
can take place, thus successively reducing the $U$-hadron abundance.
Hadronic recombination should take place even in the absence
of long range $y$-interaction of $U$-particles. So, we give first
the result without the account of radiative recombination
induced by this interaction.

There is a large uncertainties in the estimation of hadronic recombination rate.
The maximal estimation for the
reaction rate of recombination $\sv$ is given by
\beq
\sv \sim \frac{1}{m_{\pi}^2} \approx 6 \cdot 10^{-16}\,{\rm \frac{cm^3}{s}}
\label{hadsigmv}
\eeq
or by
\beq
\sv \sim \frac{1}{m_{\rho}^2} \approx 2 \cdot 10^{-17}\,{\rm \frac{cm^3}{s}}.
\label{hadsigmv1}
\eeq
The minimal realistic estimation gives \cite{BFK}
\beq \sv \approx 0.4 \cdot (T_{eff} m^3)^{-1/2}
(3 + \log{(T_{QCD}/T_{eff})}),
\label{hadrecmin}
\eeq
where $T_{eff} = \max{\{T, \alpha_y m_{\pi}\}}$.

Solution of Eq.(\ref{hadrecomb}) for $\sv$ from the Eq.(\ref{hadsigmv}) is given by

Case A
\beq
r_4 = \frac{r_0}{1 + r_0\cdot \sqrt{\frac{\pi g_{QCD}}{45}} \frac{m_{Pl}}{m_{\pi}}
\frac{T_{QCD}}{m_{\pi}} } \approx 1.0\cdot 10^{-20}
\label{hadrecsol}
\eeq
and it is $(\frac{m_{\rho}}{m_{\pi}})^2 \sim 30$ times larger for $\sv$ from
the Eq.(\ref{hadsigmv1}):

Case B
\beq
r_4 = \frac{r_0}{1 + r_0\cdot \sqrt{\frac{\pi g_{QCD}}{45}} \frac{m_{Pl}}{m_{\rho}}
\frac{T_{QCD}}{m_{\rho}} } \approx 3.0 \cdot 10^{-19}.
\label{hadrecsol2}
\eeq
For the minimal estimation of recombination rate (\ref{hadrecmin})
the solution of Eq.(\ref{hadrecomb}) has the form

\beq
r_4 = \frac{r_0}{1 + r_0\cdot 2 \cdot \sqrt{\frac{\pi
g_{QCD}}{45}} \frac{m_{Pl}}{m} \sqrt{\frac{T_{QCD}}{m}} }
\label{hadrecsol3}
\eeq
where in all the cases $r_0$ is given by
Eq.(\ref{radrecsolq}) and $g_{QCD}\approx 15$.
We neglect in our estimations possible effects of recombination in
the intermediate period, when QCD phase transition proceeds.

The solutions (\ref{hadrecsol}) and (\ref{hadrecsol2}) are
independent on the actual initial value of $r_4 = r_0$, if before
QCD phase transition it was of the order of (\ref{radrecsolq}).
For the minimal estimation of the recombination rate
(\ref{hadrecmin}) the result of hadronic recombination
 reads

Case C
\beq r_4 \approx 1.2 \cdot 10^{-16}\fr{m}{250\GeV}{3/2}.
\label{hadrecsol4}
\eeq

As we mentioned above,
for the smallest allowed mass of $U$-quark,
diquarks $(UU)$, $(\bar U \bar U)$ and the triple
$U$ (and $\bar U$) states $(UUU)$, $\anti UUU$ can
not form before QCD phase transition. Therefore U-baryonic states
$(UUu)$, $(UUU)$ and their antiparticles should
originate from single $U$ (and $\bar U$) hadron collisions. The
rate of their creation shares the same theoretical uncertainty as
in the case of $(U \bar U)$ formation, considered above. Moreover,
while baryon $(UUu)$ can be formed e.g. in reaction $(Uud) + (Uud)
\rightarrow (UUu) + n$, having no energetic threshold, formation
of antibaryon $\anti UUu$ may be suppressed at
smallest values of $m$ by the threshold of nucleon production in
reaction $(\bar U u) + (\bar U u) \rightarrow \anti UUu + p + \pi^+$,
which can even exceed $\bar U \bar U$ binding energy.
In further consideration we will not specify $\bar U$-hadronic content,
assuming that $\anti UUU$, $\anti UUu$ and $(\bar Uu)$ can be present with
appreciable fraction,
while the content of residual $U$-hadrons is likely to be realized with
multiple U-states and with suppressed fraction of single U-states.
Nevertheless we can not ignore single U-baryonic states $(Uud)^+$
because only reliable inference on their strong suppression
would avoid opposing to strong constraint on $+1$ heavy particles abundance
which will be considered below.

%\newline
\textbf{Radiative recombination}\\*
Radiative $U \bar U$ recombination is induced by "Coulomb-like"
attraction of $U$ and $\bar U$ due to their $y$-interaction. It
can be described in the analogy to the process of free
monopole-antimonopole annihilation considered in \cite{ZK}.
Potential energy of Coulomb-like interaction between $U$ and $\bar U$
exceeds their thermal energy $T$ at the distance
$$ d_0 = \frac{\alpha}{T}.$$
In the case of $y$-interaction its running constant $\alpha =
\alpha_y \sim 1/30$ \cite{Shibaev}. For $\alpha \ll 1$, on the
contrary to the case of monopoles \cite{ZK} with $g^2/4 \pi \gg
1$, the mean free path of multiple scattering in plasma is given by
$$\lambda = (n \sigma)^{-1} \sim \left(T^3 \cdot \frac{\alpha^2}{T m}\right)^{\! -1}
\sim \frac{m}{\alpha^3 T} \cdot d_0,$$
being $\lambda \gg d_0$ for all $T < m$. So the diffusion
approximation \cite{ZK} is not valid for our case. Therefore
radiative capture of free $U$ and $\bar U$ particles should be
considered. According to \cite{ZK}, following the classical
solution of energy loss due to radiation, converting infinite
motion to finite, free $U$ and $\bar U$ particles form bound
systems at the impact parameter
\beq
a \approx (T/m)^{3/10} \cdot
d_0. \label{impact}
\eeq
The rate of such binding is then given by
\beq
\sv = \pi a^2 v \approx \pi \cdot (m/T)^{9/10} \cdot \fr{\alpha}{m}2 \approx
\label{sigimpact}
\eeq
$$\approx 6 \cdot 10^{-13} \fr{\alpha}{1/30}{2} \fr{300\K}{T}{9/10}
\fr{250\GeV}{m}{11/10} {\rm\frac{cm^3}{s}}.$$

The successive evolution of this highly excited atom-like bound
system is determined by the loss of angular momentum owing to
y-radiation. The time scale for the fall on the center in this
bound system, resulting in $U \bar U$ recombination was estimated
according to classical formula in \cite{DFK}

\beq
\tau = \frac{a^3}{64 \pi} \cdot \fr{m}{\alpha}2 =
\frac{\alpha}{64 \pi} \cdot \fr{m}{T}{21/10} \cdot
\frac{1}{m} \label{recomb}
\eeq
$$\approx 4 \cdot 10^{-4} \fr{300\K}{T}{21/10} \fr{m}{250\GeV}{11/10} \s.$$

As it is easily seen from Eq.(\ref{recomb}) this time scale of $U
\bar U$ recombination $\tau \ll m/T^2 \ll m_{Pl}/T^2$ turns to be
much less than the cosmological time at which the bound system was
formed.

The above classical description assumes $a=
\frac{\alpha}{m^{3/10}T^{7/10}} \gg  \frac{1}{\alpha m}$ and is
valid at $T \ll m  \alpha^{20/7}$ \cite{Fargion:2005xz}.

Kinetic equation for U-particle abundance with the account of
radiative capture on RD stage is given by Eq.(\ref{hadrecomb}).

At $T < T_{rr} = \alpha_y^{20/7} m \approx 10 {\MeV} (m/250
\GeV) \fr{\alpha_y}{1/30}{20/7}$ the solution for
the effect of radiative recombination is given by
\beq
r_4 \approx \frac{r_0}{1 + r_0\, \sqrt{\frac{20 g_{rr}\pi^3}{9}}\,
\frac{\alpha^2 m_{Pl}}{m} \,
\left(\frac{T_{rr}}{m}\right)^{1/10}}\approx r_0
\eeq
with $r_0$ taken at $T = T_{rr}$ equal to $r_4$ from
Eqs.(\ref{hadrecsol}),(\ref{hadrecsol2}) or (\ref{hadrecsol4}).

Owing to more rapid cosmological expansion radiative capture of
$U$-hadrons in expanding matter on MD stage is less effective,
than on RD stage. So the result $r_4 \approx r_0$ holds on MD
stage with even better precision, than on RD stage. Therefore
radiative capture does not change the estimation of $U$-hadron
pregalactic abundance, given by
Eqs.(\ref{hadrecsol}),(\ref{hadrecsol2}) or (\ref{hadrecsol4}).

On the galactic stage in the most of astrophysical bodies
temperature is much less than $T_{rr}$ and radiative recombination
plays dominant role in the decrease of $U$-hadron abundance inside
dense matter bodies.

%\newline
\textbf{U-hadrons during Big Bang Nucleosynthesis and thereafter}\\*
One reminds that to the beginning of Big Bang Nucleosynthesis (BBN)
there can be $(Uud)^+$, $(UUu)^{++}$, $(UUU)^{++}$, $(\bar Uu)^0$,
$\anti UUU^{--}$, $\anti UUu^{--}$ states in plasma.
We do not specify here possible fractions of each of the U-hadron species ($i$)
in U-hadronic matter, assuming that any of them can be appreciable
($r_i\lsim r_4$).

After BBN
proceeded, the states $\anti UUU^{--}$, $\anti UUu^{--}$ are
combined with $^4He^{++}$ due to electromagnetic interaction. The
binding energy of the ground state can be estimated with reasonable
accuracy following Bohr formulas (for point-like particles)
\beq
I_{b}=\frac{(Z_AZ_X\alpha)^2m_A}{2}\approx 1.5 \MeV, \label{Ib}
\eeq
where $Z_X=2$, $Z_A=2$ and $m_A\approx 3.7$ GeV are the charges of
U-hadron and Helium and the mass of the latter.  Cross section of
this recombination is estimated as \cite{Kohri2}
\beq
\sv =
\frac{2^8\pi\sqrt{2\pi}\alpha^3Z_A^4Z^2}{3\exp(4)m_A\sqrt{m_AT}}
\approx \frac{3.06\cdot 10^{-4}}{m_A\sqrt{m_AT}}.
\label{svrec}
\eeq
Evolution of abundance of U-hadrons combining
with $He$ is described by equation
\beq
\frac{dn_{\anti
UUU}}{dt}=-3Hn_{\anti UUU}-\sv n_{\anti UUU}n_{He}.
\label{combHe}
\eeq
The term corresponding to disintegration of $[\anti UUU He]$ is
neglected, since the energy of thermal photons is insufficient to
disintegrate $[\anti UUU He]$ (the same for $[\anti UUu He]$) in the
ground state in this period. Following procedure
Eqs.(\ref{hadrecomb1}-\ref{hadrecomb}), we get
\beq
r_{\anti UUU}=r_{\anti UUU0}\exp\left({-\sqrt{\frac{\pi g}{45}}
m_{Pl}\int^{T_0}_0r_{He}\sv dT}\right)\approx,
\label{runbound}
\eeq
$$\approx r_{\anti UUU0} \exp{\left(-0.6\cdot 10^{12}\right)},$$
where $r_{He}\equiv n_{He}/s=Y_p/4 \cdot \eta\cdot
n_{\gamma}^{mod}/s_{mod}\approx 5.2\cdot 10^{-12}$, $g$ follows from
Eq.(\ref{gmod}) and $T_0=100$ keV was taken. As one can see,
Eq.(\ref{runbound}) gives in this case strong exponential
suppression of free $\anti UUU$ (the same for $\anti UUu$), while
neutral $[(\bar U \bar U \bar U)He]$ and $[(\bar U \bar U \bar
u)He]$ states, being one of the forms of $O$-helium
\cite{I,lom,KPS,FKS,Majorana,KK,Bled07,DM08},
catalyze additional annihilation of free $U$-baryons and formation
of primordial heavy elements \cite{phe}. New type of nuclear
reactions, catalyzed by $O$-helium, seem to change qualitatively the
results of BBN, however (see Sec. \ref{asymmetry} and arguments in
\cite{I,lom,KPS,FKS,Majorana,KK,Bled07,DM08,phe})
it does not lead to immediate contradiction with the observations.

On the base of existing results of investigation of hyper-nuclei \cite{hypernucl},
one can expect that the isoscalar state $\Lambda_U^+=(Uud)^+$ can form stable bound state
with $^4He$ due to nuclear interaction. The change of abundance of U-hyperons $\Lambda_U^+$
owing to their nuclear fusion with $^4He$ is described by Eq.(\ref{combHe},\ref{runbound}),
substituting $\anti UUU \leftrightarrow \Lambda_U^+$.
Disintegration of $[\Lambda_UHe]$ is also negligible, since the period, when BBN
is finished, is of interest ($T<T_0\ll I([\Lambda_UHe])$).

Cross section for nuclear reaction of question can be represented
in conventional parameterization through the so called astrophysical S-factor
\beq
\sigma = \frac{S(E)}{E}\exp\left({-\frac{2\pi \alpha Z_XZ_A}{v}}\right),
\label{sigma_Li}
\eeq
where $E=\mu v^2/2$ with $\mu$ being reduced mass of interacting particles and $v$ being their relative velocity.
The exponent in Eq.(\ref{sigma_Li}) expresses penetration factor, suppressing cross section,
which reflects repulsive character of Coulomb force contrary to the case of $\anti UUU$.
S-factor itself is unknown, being supposed $S(E\to 0)\to const$.
Averaging $\sigma v$ over Maxwell velocity distribution gives, using saddle point method,
\beq
\sv\approx \frac{4v_0\cdot S(E(v_0))}{\sqrt{3}T} \exp\left({-\frac{3\mu v_0^2}{2T}}\right),
\eeq
where $v_0=\fr{2\pi\alpha Z_AZ_XT}{\mu}{1/3}$.

Calculation gives that suppression of free $\Lambda_U^+$ on
more than 20 orders of magnitude is reached at $S(E)\gsim 2 \MeV\!\cdot\! \bn$.
S-factor for reaction of $^4He$ production is typically distinguished by high magnitudes from those
of other reactions and lies around $5-30 \MeV\!\cdot\! \bn$ \cite{crosssec}. However,
reactions with $\gamma$ in final state, which is assumed in our case
($\Lambda_U+\, ^4\!He\to [\Lambda_UHe]+\gamma$), have as a rule S-factor in $~10^4$ times smaller.
Special conditions should be demanded from unknown for sure physics of $\Lambda_U$-nucleus interaction
to provide a large suppression of $\Lambda_U$ abundance. Such suppression is needed, as we will
see below, to avoid contradiction with data on anomalous hydrogen abundance in terrestrial matter.
The experimental constraints on anomalous lithium are less restrictive and can be satisfied in this case.

%\newline
\textbf{$y$-plasma}\\*
The existence of new massless U(1) gauge boson ($y$-photon)
implies the presence of primordial thermal $y$-photon background
in the Universe. Such background should be in equilibrium with
ordinary plasma and radiation until the lightest particle bearing
$y$-charge (4th neutrino) freezes out. For the accepted value of
4th neutrino mass ($\ge 50 \GeV$) 4th neutrino freezing out and
correspondingly decoupling of $y$-photons takes place before the
QCD phase transition, when the total number of effective degrees
of freedom is sufficiently large to suppress the effects of
$y$-photon background in the period of Big Bang nucleosynthesis.
This background does not interact with nucleons and does not
influence the BBN reactions rate (its possible effect in formation
and role of $[^4He\anti UUU]$ "atom" is discussed in
\cite{BFK}), while the suppression of $y$-photon energy density
leads to insignificant effect in the speeding up cosmological
expansion rate in the BBN period. In the framework of the present
consideration the existence of primordial $y$-photons does not
play any significant role in the successive evolution of
$U$-hadrons.

Inclusion of stable $y$-charged 4th neutrinos strongly complicate the picture.
Condition of cancellation of axial anomalies
requires relationship between the values of $y$-charges of 4th generation
leptons ($N,E$) and quarks ($U,D$) as the following
$$e_{yN}=e_{yE}=-e_{yU}/3=-e_{yD}/3.$$
In course of cosmological combined evolution of $U$ and $N$ and $y$, ``$y$-molecules''
of kind U-U-U-N, where different U-quarks can belong to different U-hadrons
(possibly bound with nucleus) should form. Such $y$-neutral molecules can avoid effect of
U-hadrons suppression in the terrestrial matter,
relevant in charge symmetric case,
and lead to contradiction with observations,
analysis of which is started now.
UUU-N-type states will be considered in section \ref{asymmetry} devoted to the charge-asymmetric case.

%%%%%%%%%%%%%%%%%%%%%%%%%%%%%%%%%%%%%%%%%%%%%%%%%%%%%%%%%%%%%%%%%%%%%%%%%%%%%%%%%%%
\subsection{\label{matter} Evolution and manifestations of $U$-hadrons at galactic stage}

In the period of recombination of nuclei with electrons
the positively charged $U$-baryons
recombine with electrons to form
atoms of anomalous isotopes. The substantial (up to 10 orders
of magnitude) excess of electron number density over the number density
of primordial $U$-baryons makes virtually all $U$-baryons to form atoms.
The cosmological abundance of free charged $U$-baryons is to be exponentially
small after recombination.

Hadrons $(UUu)$, $(UUU)$ form atoms of anomalous He
at $T \sim 2 \eV$ together with recombination of ordinary helium.
The states $[\anti UUU He]$, $[\anti UUu He]$,
$(\bar Uu)$ escape recombination with electrons because of their neutrality; hadrons $(Uud)$,
if they are not involved into chain of nuclear transitions,
form atoms of anomalous hydrogen.

The formed atoms, having atomic cross sections of interaction with
matter follow baryonic matter in formation of
astrophysical objects like gas clouds, stars and planets,
when galaxies are formed.

On the contrary, O-helium and $(\bar U u)$ mesons, having nuclear
and hadronic cross sections, respectively, can decouple from plasma
and radiation at $T \sim 1 \keV$ and behave in Galaxy as
collisionless gas. In charge asymmetric case, considered in the next
Section \ref{asymmetry}, or in charge symmetric case without
$y$-interaction O-helium and $(\bar U u)$ mesons behave on this
reason as collisionless gas of dark matter particles.
On that reasons one can expect suppression of their
concentration in baryonic matter.

However, in charge symmetric case with $y$-interaction,
the existence of Coulomb-like $y$-attraction will make them to obey the condition
of neutrality in respect to the $y$-charge. Therefore owing to
neutrality condition the number densities of $U$- and $\bar
U$-hadrons in astrophysical bodies should be equal.
It leads to effects in
matter bodies, considered in this subsection.

\textbf{\indent U-hadrons in galactic matter}\\*
In the
astrophysical body with atomic number density $n_a$ the initial
$U$-hadron abundance $n_{U0} = f_{a0} \cdot n_a$ can decrease with
time due to $U \bar U$ recombination.
Here and in estimations thereafter we will refer to U-quark abundance as
U-hadron one (as if all U-hadrons were composed of single U-quarks),
if it is not specified otherwise.

Under the neutrality condition
$$n_U = n_{\bar U}$$
the relative $U$-hadron abundance $f_{a0} = n_U/n_a= n_{\bar U}/n_a$ is governed by the equation
\beq
\frac{df_a}{dt} = - f_a^2 \cdot n_a \cdot \sv.
\label{recomb}
\eeq
Here $\sv$ is defined by Eq.(\ref{sigimpact}). The solution of this equation is given by
\beq
f_a = \frac{f_{a0}}{1 + f_{a0} \cdot n_a \cdot \sv \cdot t}.
\label{sol}
\eeq
If
\beq
n_a \cdot \sv \cdot t \gg \frac{1}{f_{a0}},
\label{cond}
\eeq
the solution (\ref{sol}) takes the form
\beq
f_a = \frac{1}{n_a \cdot \sv \cdot t}.
\label{sol2}
\eeq
and, being independent on the initial value, $U$-hadron abundance
decreases inversely proportional to time.

By definition $f_{a0}= f_0/A_{atom}$, where $A_{atom}$ is the averaged atomic weight of the considered matter
and $f_0$ is the initial $U$-hadron to baryon ratio.
In the pregalactic matter this ratio is determined by $r_4$ from A) Eq.(\ref{hadrecsol}), B)
Eq.(\ref{hadrecsol2}) and C) Eq.(\ref{hadrecsol4}) and is
equal to
\beq
f = \frac{r_4}{r_b} =\left\{
 \begin{array}{c}
 10^{-10} \;\; {\rm for \, the \, case \, A},\\
 3 \cdot 10^{-9} \;\; {\rm for \, the \, case \,B},\\
 1.2 \cdot  10^{-6} \;\; {\rm for \, the \, case \,C}.
 \end{array}
\right.
\label{primcr1}
\eeq
Here $r_b \approx 10^{-10}$ is baryon to entropy ratio.

Taking for averaged atomic number density in the Earth $n_a \approx 10^{23}
\cm^{-3}$, one finds that during the age of the Solar system
primordial $U$-hadron abundance in the terrestrial matter should
have been reduced down to $f_a \approx 10^{-28}$. One should
expect similar reduction of $U$-hadron concentration in Sun and
all the other old sufficiently dense astrophysical bodies.
Therefore in our own body we might contain just one of such heavy hadrons.
However, as shown later on,
the persistent pollution
from the galactic gas nevertheless may increase this relic number density
to much larger value ($f_a \approx 10^{-23}$).

The principal possibility of strong reduction in dense bodies for
primordial abundance of exotic charge symmetric particles
due to their recombination in unstable charmonium like systems
was first revealed in \cite{fractons} for fractionally charged
colorless composite particles (fractons).

The $U$-hadron abundance in the interstellar gas strongly depends
on the matter evolution in Galaxy, which is still not known to the extent,
we need for our discussion.

Indeed, in the opposite case of low density or of short time interval,
when the condition (\ref{cond}) is not valid, namely, at
\beq
n_a < \frac{1}{f_{a0} \sv t} = A_{atom} \cdot \frac{T}{300\K} \cdot \frac{t_U}{t} \cm^{-3} \left\{
 \begin{array}{c}
 4 \cdot 10^{4} \; {\rm for \, the \, case \, A},\\
 10^{3} \; {\rm for \, the \, case \, B},\\
1.2 \; {\rm for \, the \, case \, C},
 \end{array}
\right.
\label{lowcond}
\eeq
where $t_U=4 \cdot 10^{17}$ s is the age of the Universe,
 $U$-hadron abundance does not change its initial value.

In principle, if in the course of evolution matter in the forming
Galaxy was present during sufficiently long period ($t \sim 10^9 $
yrs) within cold ($T \sim 10$ K) clouds with density $n_a \sim 10^3
\cm^{-3}$ $U$-hadron abundance retains its primordial value for the
cases A and B ($f_0 = f$), but falls down $f_0 = 5 \cdot 10^{-9}$ in
the case C, making this case close to the case B. The above argument
may not imply all the $U$-hadrons to be initially present in cold
clouds. They can pass through cold clouds and decrease their
abundance in the case C at the stage of thermal instability, when
cooling gas clouds, before they become gravitationally bound, are
bound by the external pressure of the hot gas. Owing to their large
inertia heavy $U$-hadron atoms from the hot gas can penetrate  much
deeply inside the cloud and can be captured by it much more
effectively, than ordinary atoms. Such mechanism can provide
additional support for reduction of $U$-hadron abundance in the case
C. The reduction of this abundance down to values,
corresponding to the case A can be also provided by $O$-helium
catalysis in the period of BBN.

However, in particular, annihilation of $U$-hadrons leads to multiple
$\gamma$ production. If $U$-hadrons with relative abundance $f$
annihilate at the redshift $z$, it should leave in the modern
Universe a background $\gamma$ flux \cite{BFK}
$$F(E>E_{\gamma}) = \frac{N_{\gamma} \cdot f \cdot r_b \cdot s_{mod} \cdot c}{4 \pi}
\approx 3 \cdot 10^3 f \,{\rm ( cm^2 \cdot s \cdot ster)^{-1}},$$
of $\gamma$ quanta with energies $E>E_{\gamma} = 10 \GeV/(1+z)$.
The numerical values for $\gamma$ multiplicity $N_{\gamma}$
are given in table 1 \cite{BFK}.

\begin{table}[h]
 \begin{tabular}[]{|c|c|c|c|c|c|}
\hline
{\small
Energy fraction} & $>0$ & $>0.1$ GeV & $>1$ GeV & $>10$ GeV & $>100$ GeV \\ \hline
$N_{\gamma}$ & 69 & 62 & 28 & 2.4 & 0.001  \\ \hline
\end{tabular}
\caption{Multiplicities of $\gamma$ produced in the recombination of
$(Q \bar Q)$ pair with $m=250$ GeV for different energy intervals.}
\end{table}

So annihilation even as early as at $z \sim 9$
leads in the case C to the contribution into diffuse extragalactic gamma emission,
exceeding the flux, measured by EGRET by three orders of magnitude.
The latter can be approximated as
$$F(E>E_{\gamma}) \approx 3\cdot 10^{-6} \fr{E_0}{E_{\gamma}}{1.1} ({\rm cm^2 \cdot s \cdot ster})^{-1},$$
where $E_0 = 451 \MeV$.

The above upper bound strongly restricts ($f \le 10^{-9}$) the earliest abundance
because of the consequent impossibility to reduce the primordial $U$-hadron abundance
by $U$-hadron annihilation in low density objects. In the cases A and B annihilation in
such objects should not take place, whereas annihilation within the dense objects,
being opaque for $\gamma$ radiation, can avoid this constraint due to strong suppression of outgoing
$\gamma$ flux. However, such constraint nevertheless should arise for the period of dense objects' formation.
For example, in the course of protostellar collapse hydrodynamical timescale
$t_H \sim 1/\sqrt{\pi G \rho} \sim 10^{15} \s /\sqrt{n}$
exceeds the annihilation timescale \cite{BFK}
$$t_{an} \sim \frac{1}{f n \sv} \sim  \frac{10^{12} \s}{f n}
\fr{1/30}{\alpha}{2} \fr{T}{300\K}{9/10} \fr{m}{250\GeV}{11/10}$$
at $n > 10^{14} \fr{10^{-10}}{f}{2} \fr{1/30}{\alpha}{4} \fr{T}{300\K}{9/5} \fr{m}{250\GeV}{11/5}$,
where $n$ is in cm$^{-3}$. We consider homogeneous cloud with mass $M$ has radius
$R \approx \frac{10^{19}\cm}{n^{1/3}} \fr{M}{M_{\odot}}{1/3}$,
where $ M_{\odot} = 2 \cdot 10^{33}$ g is the Solar mass. It becomes opaque for $\gamma$ radiation, when this radius
exceeds the mean free path $l_{\gamma} \sim 10^{26} \cm/n$ at $n> 3 \cdot 10^{10}\cm^{-3} \fr{M}{M_{\odot}}{1/2}$.
As a result, for $f$ as large as in the case C, rapid annihilation takes place when the collapsing matter is
transparent for $\gamma$ radiation and the EGRET constraint can not be avoided. The cases A and B are consistent with this constraint.

Note that at $f < 5 \cdot 10^{-6} \fr{M}{M_{\odot}}{2/9}$, i.e. for
all the considered cases energy release from $U$-hadron annihilation
does not exceed the gravitational binding energy of the collapsing
body. Therefore, $U$-hadron annihilation can not prevent the
formation of dense objects but it can provide additional energy
source, e.g. at early stages of evolution of first stars. Its
burning is quite fast (few years) and its luminosity may be quite
extreme, leading to a short inhibition of star formation \cite{BFK}.
Similar effects of dark matter annihilation were recently
considered in \cite{DMstars}.

%\newline
\textbf{Galactic blowing of $U$-baryon atoms polluting our Earth}\\*
Since the condition Eq.(\ref{lowcond}) is valid for the disc
interstellar gas, having the number density $n_g \sim 1\cm^{-3}$ one can expect
that the $U$-hadron abundance in it can decrease relative to the
primordial value only due to enrichment of this gas by the matter,
which has passed through stars and had the
suppressed $U$-hadron abundance according to Eq.(\ref{sol2}).
Taking the factor of such decrease of the order of the
ratio of total masses of gas and stars in Galaxy $f_g \sim 10^{-2}$
and accounting for the acceleration of the interstellar gas
by Solar gravitational force, so that the infalling gas has velocity
$v_g \approx 4.2 \cdot 10^6$cm/s in vicinity of Earth's orbit,
one obtains that the flux of $U$-hadrons coming with interstellar
gas should be of the order of \cite{BFK}
\beq
I_U = \frac{f f_g n_g v_g}{8 \pi} \approx 1.5 \cdot 10^{-7} \frac{f}{10^{-10}}\,{\rm (sm^2\cdot s \cdot ster)}^{-1},
\label{interst}
\eeq
where $f$ is given by the Eq.(\ref{primcr1}).

Presence of primordial $U$-hadrons in the Universe should be reflected by their existence
in Earth's atmosphere and ground.
However, according to  Eq.(\ref{sol2}) (see
discussion in Section \ref{matter}) primordial
terrestrial $U$-hadron content should strongly decrease due to
radiative recombination, so that the $U$-hadron
abundance in Earth is determined by the kinetic
equilibrium between the incoming $U$-hadron flux
and the rate of decrease of this abundance
by different mechanisms.

In the successive analysis we'll concentrate our attention on the case,
when the $U$ baryon has charge $+2$, and $\bar U$-hadrons are
electrically neutral.
In this case $U$ baryons look like superheavy anomalous helium
isotopes.

Searches for anomalous helium were performed in series of
experiments based on accelerator search \cite{exp1}, spectrometry
technique \cite{exp2} and laser spectroscopy \cite{exp3}.
From the experimental point of view an anomalous helium
represents a favorable case, since it remains in the atmosphere
whereas a normal helium is severely depleted in the terrestrial
environment due to its light mass.

The best
upper limits on the anomalous helium were obtain in \cite{exp3}.
It was found by searching for a heavy helium
isotope in the Earth's atmosphere that in the mass range 5 GeV -- 10000 GeV the
terrestrial abundance (the ratio of anomalous helium number to the
total number of atoms in the Earth) of anomalous helium is less
than $(2 - 3) \cdot 10^{-19}$. The search in the atmosphere is reasonable because heavy
gases are well mixed up to 80 km and because the heavy helium
does not sink due to gravity deeply in the Earth and is
homogeneously redistributed in the volume of the World Ocean at
the timescale of $10^3$ yr.

The kinetic equations, describing evolution of anomalous helium and $\bar U$-hadrons
in matter have the form \cite{BFK}
\beq
\frac{dn_{U}}{dt} = j_{U} - n_{U} \cdot n_{\bar U} \cdot \sv - j_{gU}
\label{kin1}
\eeq
for  $\bar U$-hadron number density $n$ and
\beq
\frac{dn_{\bar U}}{dt} = j_{\bar U} - n_{\bar U} \cdot n_{U} \cdot \sv - j_{g \bar U}
\label{kin2}
\eeq
for number density of anomalous helium $n_U$.
Here $j_{U}$ and $j_{\bar U}$ take into account the income of, correspondingly, $U$-baryons
and $\bar U$-hadrons to considered region, the second terms on the right-hand-side of equations
describe $U \bar U$ recombination and the terms $j_{gU}$ and $j_{g \bar U}$ determine
various mechanisms for outgoing fluxes, e.g. gravitationally driven sink of particles.
The latter effect is much stronger for $\bar U$-hadrons due to much lower mobility
of $U$-baryon atoms. However, long range Coulomb like interaction prevents them from
sinking, provided that its force exceeds the Earth's gravitational force.

In order to compare these forces let's consider the World's Ocean as a thin shell
of thickness $L \approx 4 \cdot 10^5$ cm
with homogeneously distributed $y$ charge,
determined by distribution of $U$-baryon atoms with concentration $n$. The
$y$-field outside this shell according to Gauss' law is determined by
$$2 E_y S = 4 \pi e_y n S L,$$
being equal to
$$E_y = 2 \pi e_y n L.$$
In the result $y$ force, exerting on $\bar U$-hadrons
$$F_y = e_y E_y,$$  exceeds gravitational force for $U$-baryon atom concentration

\beq
n > 10^{-7} \frac{m}{250\GeV} \frac{30^{-1}}{\alpha_y} \cm^{-3}.
\label{neutr}
\eeq

Note that the mobility of $U$-baryon atoms
and $\bar U$ hadrons differs by 10 order of magnitude,
what can lead to appearance of excessive $y$-charges
within the limits of (\ref{neutr}). One can expect
that such excessive charges arise due to the effective slowing down
of $U$-baryon atoms in high altitude levels
of Earth's atmosphere,
which are transparent for $\bar U$ hadrons,
as well as due to the 3 order of magnitude decrease of $\bar U$ hadrons
when they enter the Earth's surface.

Under the condition of neutrality, which is strongly protected by
Coulomb-like $y$-interaction, all the corresponding parameters for $\bar U$-hadrons
and $U$-baryons in the Eqs.(\ref{kin1})-(\ref{kin2}) are equal, if Eq.(\ref{neutr}) is valid.
Provided that the timescale of mass exchange between the Ocean and atmosphere is much less than
the timescale of sinking, sink terms can be neglected.

The stationary solution of Eqs.(\ref{kin1})-(\ref{kin2}) gives in this case
\beq
n = \sqrt{\frac{j}{\sv}},
\label{statsol}
\eeq
where
\beq
j_{U}=j_{\bar U}=j \sim \frac{2 \pi I_U}{L} = 10^{-12} \frac{f}{10^{-10}}\, {\rm cm^{-3}s^{-1}}
\label{statin}
\eeq
and $\sv$ is given by the Eq.(\ref{sigimpact}).
For $j \le 10^{-12} \frac{f}{10^{-10}}\, {\rm cm^{-3}s^{-1}}$
and $\sv$ given by Eq.(\ref{sigimpact}) one obtains in water
$$ n \le \sqrt{\frac{f}{10^{-10}}} \cm^{-3}.$$
It corresponds to terrestrial $U$-baryon abundance
$$ f_a \le 10^{-23}\sqrt{\frac{f}{10^{-10}}},$$
being below the above mentioned experimental upper limits for anomalous helium ($ f_a < 10^{-19}$)
even for the case C with $f= 2\cdot 10^{-6}$.
In air one has
$$ n \le 10^{-3}\sqrt{\frac{f}{10^{-10}}} \cm^{-3}.$$
For example in a cubic room of 3m size there are nearly 27 thousand heavy hadrons.

Note that putting formally in the Eq.(\ref{statsol}) the value of $\sv$
given by the Eq.(\ref{hadsigmv}) one obtains $ n \le 6 \cdot 10^2 \cm^{-3}$
and $ f_a \le 6 \cdot 10^{-20}$, being still below the experimental upper limits
for anomalous helium abundance. So the qualitative conclusion that
recombination in dense matter can provide the sufficient decrease
of this abundance avoiding the contradiction with the experimental constraints
could be valid even in the absence of gauge $y$-charge and Coulomb-like $y$-field
interaction for $U$-hadrons. It looks like the hadronic recombination
alone can be sufficiently effective in such decrease. However, if we take
the value of $\sv$
given by the Eq.(\ref{hadsigmv1}) one obtains $n$ by the factor of $\frac{m_{\rho}}{m_{\pi}} \sim 5.5$
larger and $ f_a \le 3.3 \cdot 10^{-19}$, what exceeds the experimental upper limits
for anomalous helium abundance.
Moreover, in the absence
of $y$-attraction there is no dynamical mechanism, making the number densities
of $U$-baryons and $\bar U$-hadrons equal each other with high accuracy.
So nothing seems to prevent in this case selecting and segregating
$U$-baryons from $\bar U$-hadrons. Such segregation,
being highly probable due to the large difference in the mobility of $U$-baryon atoms
and $\bar U$ hadrons can lead to uncompensated excess of anomalous helium in the Earth,
coming into contradiction with the experimental constraints.

Similar result can be obtained for any planet, having atmosphere and Ocean,
in which effective mass exchange between atmosphere and Ocean takes place.
There is no such mass exchange in planets without atmosphere and Ocean
(e.g. in Moon) and $U$-hadron abundance in such planets is determined
by the interplay of effects of incoming interstellar gas, $U \bar U$ recombination
and slow sinking of $U$-hadrons to the centers of planets.

Radiative recombination here considered is not able to save the case
when free single positively charged U-hadrons $(Uud)$ are present with
noticeable fraction among U-hadronic relics. It is the very strong
constraint on anomalous hydrogen ($f_a<10^{-30}$) \cite{anomH}
what is hardly avoidable.
So, the model with free stable $+1$ charged relics which are not bound
in nuclear systems with charge $\ge +2$
can be discarded.

\begin{figure}
\includegraphics[scale=0.6]{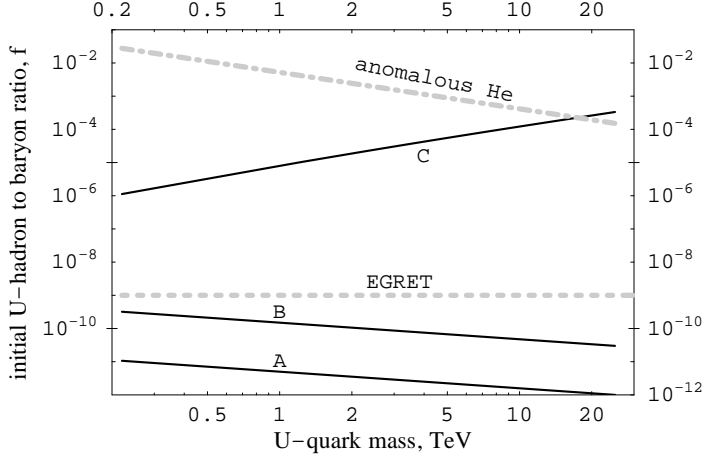}
\caption{Upper constraints on relative abundances of U-hadrons ($f$)
following from the search for anomalous helium in the Earth,
from EGRET, constraining possible annihilation effect on pregalactic stage,
in comparison with prediction $f$ in three cases (A,B,C).}
\label{constraints}
\end{figure}

To conclude this section we present the obtained constraints on the Fig.(\ref{constraints}).

%\newline
\textbf{Cosmic rays experiments}\\*
In addition to the possible traces of U-hadrons existence in the Earth, they can be manifested in cosmic rays.

According to the arguments in previous section $U$-baryon abundance
in the primary cosmic rays can be close to the primordial value $f$.
It gives for case B
\beq
f = \frac{r_4}{r_b} \sim 3 \cdot 10^{-9}.
\label{primcr}
\eeq
If $U$-baryons have mostly the form $(UUU)$, its
fraction in cosmic ray helium component can reach in this case the value
$$\frac{(UUU)}{^4 He} \sim 3 \cdot 10^{-8},$$
which is accessible for cosmic rays experiments,
such as RIM-PAMELA, being under run, and AMS 02 on International Space Station.

\begin{figure}
\includegraphics[scale=0.7]{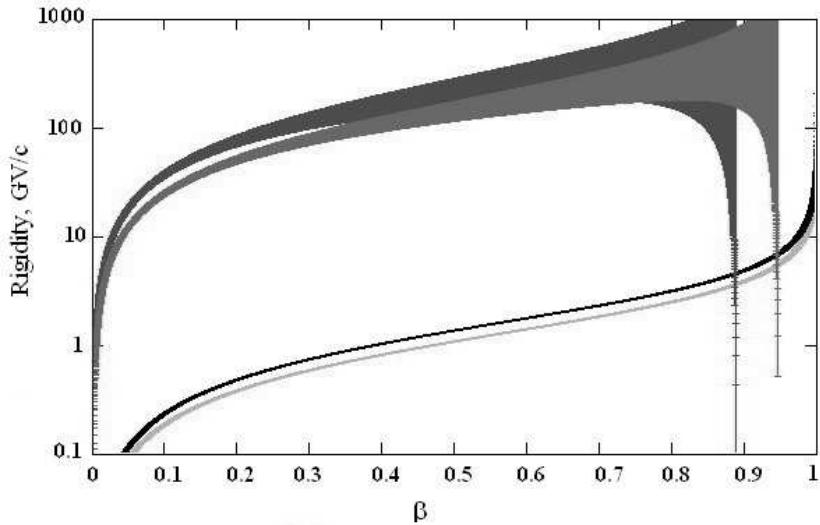}
\caption{Lines show the region of expected signal, their thickness reflects expected accuricy in experiment PAMELA.
Upper thick curves relate to anomalous helium of 500 and 750 GeV (a little upper and darker). The low thin curves
relate to usual nuclei.}
\label{PAMELA1}
\end{figure}

Similar argument in the case C would give for this fraction $\sim 2 \cdot 10^{-5}$,
what may be already excluded by the existing data. However,
it should be noted that the above estimation assumes significant
contribution of particles from interstellar matter to
cosmic rays. If cosmic ray particles are dominantly
originated from the purely stellar matter, the decrease of
$U$-hadron abundance in stars would substantially reduce
the primary $U$-baryon fraction of cosmic rays.
But in cosmic rays there are many different kind of particles.
In order to differ U-hadrons from background events
we can use the dependence of rigidity on velocity given on Fig.\ref{PAMELA1}.

The expected signal will strongly differ from background events in terms of
relation between velocity and rigidity (momentum related to the charge).
In the experiment PAMELA velocity is measured with a good accuracy, what
can lead to the picture Fig.\ref{PAMELA1}.

%\newline
\textbf{Correlation between cosmic ray and large volume underground detectors' effects}\\*
Inside large volume underground detectors (as Super Kamiokande) and in their vicinity $U$-hadron recombination
should cause specific events ("spherical" energy release with zero total momentum or "wide cone" energy release
with small total momentum), which could be clearly distinguished from the (energy release with high
total momentum within "narrow cone") effects of common atmospheric neutrino - nucleon-lepton chain
(as well as of hypothetical WIMP annihilation in Sun and Earth) \cite{BFK}.

The absence of such events
inside 22 kilotons of water in Super Kamiokande (SK) detector during 5 years of its operation
would give the most severe constraint
$$n < 10^{-3} \cm^{-3},$$
corresponding to $f_a < 10^{-26}$. For the considered type of anomalous helium such constraint would be
by 7 order of magnitude stronger, than the results of present direct searches and 3 orders above our
estimation in previous Section.

However, this constraint assumes that distilled water in SK does still contain
polluted heavy hadrons (as it may be untrue). Nevertheless even for pure water
it may not be the case for
the detector's container and its vicinity. The conservative limit follows from the condition
that the rate of $U$-hadron recombination in the body of detector does not exceed
the rate of processes, induced by atmospheric muons and neutrinos.
The  presence of clustered-like muons originated on the SK walls
would be probably observed.

High sensitivity of large volume detectors to the effects of $U$-hadron recombination
together with the expected increase of volumes of such detectors up to 1 km$^3$
offer the possibility of correlated search for cosmic ray $U$-hadrons and
for effects of their recombination.

During one year of operation a 1 km$^3$ detector could be sensitive to effects of recombination
at the $U$-hadron number density $n \approx 7 \cdot 10^{-6} \cm^{-3}$ and $f_a \approx 7 \cdot 10^{-29}$,
covering the whole possible range of these parameters, since this level of sensitivity
corresponds to the residual concentration of primordial $U$-hadrons, which can survive
inside the Earth. The income of cosmic $U$-hadrons and equilibrium between this income
and recombination should lead to increase of effect,
expected in large volume detectors.

Even, if the income of anomalous helium with interstellar gas is completely suppressed,
pollution of Earth by $U$-hadrons from primary cosmic rays is possible.
The minimal effect of pollution by $U$-hadron primary cosmic rays flux $I_U$ corresponds to the
rate of increase of $U$-hadron number density $j \sim \frac{2 \pi I_U}{R_E}$, where $R_E \approx 6 \cdot 10^8 \cm$
is the Earth's radius. If incoming cosmic rays doubly charged
$U$-baryons after their slowing down in matter
recombine with electrons we should take instead of $R_E$
the Ocean's thickness $L \approx 4 \cdot 10^5 \cm$
that increases by 3 orders of magnitude the minimal flux and the minimal number of events, estimated below.
Equilibrium between this income rate and the
rate of recombination should lead to $N \sim jVt$ events of recombination
inside the detector with volume $V$ during its operation time $t$.

For the minimal flux of cosmic ray $U$-hadrons, accessible to AMS 02 experiment
during 3 years of its operation ($I_{min} \sim 10^{-9} I_{\alpha} \sim 4 \cdot 10^{-11} I(E)$,
in the range of energy per nucleon $1 < E < 10 \GeV$)
the minimal number of events expected in detector of volume $V$ during time $t$
is given by $N_{min} \sim \frac{2 \pi I_{min}}{R_E}Vt$. It gives about 3 events
per 10 years in SuperKamiokande ($V=2.2 \cdot 10^{10} \cm^3$) and about $10^4$ events in the 1 km$^3$ detector
during one year of its operation. The noise of this rate is one order and half below
the expected influence of atmospheric $\nu_{\mu}$.

The possibility of such correlation facilitates the search for
anomalous helium in cosmic rays and for the effects of $U$-hadron recombination
in the large volume detectors.

The previous discussion assumed the lifetime of $U$-quarks $\tau$ exceeding
the age of the Universe $t_U$. In the opposite case $\tau < t_U$ all the primordial
$U$ hadrons should decay to the present time and the cosmic ray interaction
may be the only source of cosmic and terrestrial $U$ hadrons.

%%%%%%%%%%%%%%%%%%%%%%%%%%%%%%%%%%%%%%%%%%%%%%%%%%%%%%%%%%%%%%%%%%%%%%%%%%%%%%%%%%%
%%%%%%%%%%%%%%%%%%%%%%%%%%%%%%%%%%%%%%%%%%%%%%%%%%%%%%%%%%%%%%%%%%%%%%%%%%%%%%%%%%%
\section{\label{asymmetry} The case of a charge-asymmetry of U-quarks}

The model \cite{BFK} admits that in the early Universe an antibaryon
asymmetry for 4th generation quarks can be generated
\cite{I,lom,KPS}. Due to $y$-charge conservation $\bar U$ excess
should be compensated by $\bar N$ excess. We will focus our
attention here to the case of $y$-charged quarks and neutrinos of
4th generation and follow \cite{KPS} in our discussion. All
the main results concerning observational effects, presented here,
can be generalized for the case without $y$-interaction.

$\bar U$-antibaryon
density can be expressed through the modern dark matter density
$\Omega_{\bar U}= k \cdot \Omega_{CDM}=0.224$ ($k \le 1$),
saturating it at $k=1$.
It is convenient
to relate the baryon (corresponding to $\Omega_b=0.044$) and $\bar U$ ($\bar
N$) excess with the entropy density $s$, introducing $r_b = n_b/s$
and $r_{\bar U}=n_{\bar U}/s=3 \cdot n_{\bar N}/s=3 \cdot r_{\bar
N}$. One obtains $r_b \sim 8 \cdot 10^{-11}$ and $r_{\bar U},$
corresponding to $\bar U$ excess in the early Universe
$\kappa_{\bar U} =r_{\bar U} -r_{U}= 3 \cdot (r_{\bar N}
-r_{N})=10^{-12} (350 \GeV/m_U) = 10^{-12}/S_5,$ where $S_5 =
m_U/350{\GeV}$.

%\newline
\textbf{Primordial composite forms of 4th generation dark matter}\\*
In the early Universe at temperatures highly above their masses
$\bar U$ and $\bar N$ were in thermodynamical equilibrium with
relativistic plasma. It means that at $T>m_U$ ($T>m_N$) the
excessive $\bar U$ ($\bar N$) were accompanied by $U \bar U$ ($N
\bar N$) pairs.

Due to $\bar U$ excess frozen out concentration of deficit
$U$-quarks is suppressed at $T<m_U$ for $k>0.04$ \cite{lom}. It
decreases further exponentially first at $T \sim I_U \approx \bar
\alpha^2 M_U/2 \sim 3  S_5$GeV (where \cite{BFK} $\bar \alpha= C_F
\alpha_{c} = 4/3 \cdot 0.144 \approx 0.19$ and $M_U = m_U/2$ is
the reduced mass), when the frozen out $U$ quarks begin to bind
with antiquarks $\bar U $ into charmonium-like state $(\bar U U)$
and annihilate. On this line $\bar U$ excess binds at $T < I_U$ by
chromo-Coulomb forces dominantly into $\anti UUU$
anutium states with electric charge $Z_{\Delta}=-2$ and mass
$m_o=1.05 S_5$TeV, while remaining free $\bar U$ anti-quarks and
anti-diquarks $(\bar U \bar U)$ form after QCD phase transition
normal size hadrons $(\bar U u)$ and $(\bar U \bar U \bar u)$.
Then at $T = T_{QCD} \approx 150$MeV additional suppression of
remaining $U$-quark hadrons takes place in their hadronic
collisions with $\bar U$-hadrons, in which $(\bar U U)$ states are
formed and $U$-quarks successively annihilate.

Effect of $\bar N$ excess in the
suppression of deficit N takes place at
$T<m_N$ for $k>0.02$ \cite{lom}. At $T \sim I_{NN} =
\alpha_y^2 M_N/4 \sim 15 $MeV (for $\alpha_y = 1/30$ and
$M_N=50$GeV) due to $y$-interaction the frozen out $N$ begin to
bind with $\bar N $ into charmonium-like states $(\bar N N)$ and
annihilate. At $T < I_{NU} = \alpha_y^2 M_N/2 \sim 30 $MeV
$y$-interaction causes binding of $N$ with $\bar U$-hadrons
(dominantly with anutium) but only at $T \sim I_{NU}/30 \sim 1$MeV
this binding is not prevented by back reaction of
$y$-photo-destruction.

To the period of Standard Big Bang Nucleosynthesis (SBBN) $\bar U$
are dominantly bound in anutium $\Delta^{--}_{3 \bar U}$ with small
fraction ($\sim 10^{-6}$) of neutral $(\bar U u)$ and doubly charged
$(\bar U \bar U \bar u)$ hadron states. The dominant fraction of
anutium is bound by $y$-interaction with $\bar N$ in $(\bar N
\Delta^{--}_{3 \bar U})$ "atomic" state. Owing to early decoupling
of $y$-photons from relativistic plasma presence of $y$-radiation
background does not influence SBBN processes \cite{BFK,I,KPS}.

At $T<I_{o} = Z^2 Z_{He}^2 \alpha^2 m_{He}/2 \approx 1.6$MeV the
reaction $\Delta^{--}_{3 \bar U}+^4He\rightarrow \gamma
+(^4He^{++}\Delta^{--}_{3 \bar U})$ might take place, but it can
go only after $^4He$ is formed in SBBN at $T<100 $keV and is
effective only at $T \le T_{rHe} \sim
I_{o}/\log{\left(n_{\gamma}/n_{He}\right)} \approx I_{o}/27
\approx 60 $keV, when the inverse reaction of photo-destruction
cannot prevent it \cite{Fargion:2005xz,FKS,I,Khlopov:2006uv}. In
this period anutium is dominantly bound with $\bar N$. Since
$r_{He}=0.1 r_{b} \gg r_{\Delta}= r_{\bar U}/3 $, in this reaction
all free negatively charged particles are bound with helium
\cite{Fargion:2005xz,FKS,I,Khlopov:2006uv} and neutral
Anti-Neutrino-O-helium (ANO-helium, $ANOHe$) $(^4He^{++} [\bar N
\Delta^{--}_{3 \bar U}])$ ``molecule'' is produced with mass
$m_{OHe} \approx m_o \approx 1S_5$TeV. The size of this
``molecule'' is $ R_{o} \sim 1/(Z_{\Delta} Z_{He}\alpha m_{He})
\approx 2 \cdot 10^{-13}$ cm
 and it can play the role of a dark matter component and
a nontrivial catalyzing role in nuclear transformations.

In nuclear processes ANO-helium looks like an $\alpha$ particle
with shielded electric charge. It can closely approach nuclei due
to the absence of a Coulomb barrier and opens the way to form
heavy nuclei in SBBN. This path of nuclear transformations
involves the fraction of baryons not exceeding $10^{-7}$ \cite{I}
and it can not be excluded by observations.

%\newline
\textbf{ANO-helium catalyzed processes}\\*
As soon as ANO-helium is formed, it catalyzes annihilation of
deficit $U$-hadrons and $N$. Charged $U$-hadrons penetrate neutral
ANO-helium, expel $^4He$, bind with anutium and annihilate falling
down the center of this bound system. The rate of this reaction is
$\sv= \pi R^2_o$ and an $\bar U$ excess $k=10^{-3}$ is sufficient
to reduce the primordial abundance of $(Uud)$ below the
experimental upper limits. $N$ capture rate is determined by the
size of $(\bar N \Delta)$ "atom" in ANO-helium and its
annihilation is less effective.

The size of ANO-helium is of the order of the size of $^4He$ and
for a nucleus A with electric charge $Z>2$ the size of the Bohr
orbit for a $(Z \Delta)$ ion is less than the size of nucleus A.
This means that while binding with a heavy nucleus $\Delta$
penetrates it and effectively interacts with a part of the nucleus
with a size less than the corresponding Bohr orbit. This size
corresponds to the size of $^4He$, making O-helium the most bound
$(Z \Delta)$-atomic state.

The cross section for $\Delta$ interaction with hadrons is
suppressed by factor $\sim (p_h/p_{\Delta})^2 \sim
(r_{\Delta}/r_h)^2 \approx 10^{-4}/S_5^2$, where $p_h$ and
$p_{\Delta}$ are quark transverse momenta in normal hadrons and in
anutium, respectively. Therefore anutium component of $(ANOHe)$
can hardly be captured and bound with nucleus due to strong
interaction. However, interaction of the $^4He$ component of
$(ANOHe)$ with a $^A_ZQ$ nucleus can lead to a nuclear
transformation due to the reaction $^A_ZQ+(\Delta He) \rightarrow
^{A+4}_{Z+2}Q +\Delta,$ provided that the masses of the initial
and final nuclei satisfy the energy condition $M(A,Z) + M(4,2) -
I_{o}> M(A+4,Z+2),$ where $I_{o} = 1.6$MeV is the binding energy
of O-helium and $M(4,2)$ is the mass of the $^4He$ nucleus. The
final nucleus is formed in the excited $[\alpha, M(A,Z)]$ state,
which can rapidly experience $\alpha$- decay, giving rise to
$(ANOHe)$ regeneration and to effective quasi-elastic process of
$(ANOHe)$-nucleus scattering. It leads to possible suppression of
ANO-helium catalysis of nuclear transformations in matter.

%\newline
\textbf{ANO-helium dark matter}\\*
At $T < T_{od} \approx 1 \keV$ energy and momentum transfer from
baryons to ANO-helium $n_b \sv (m_p/m_o) t < 1$ is not effective.
Here $\sigma \approx \sigma_{o} \sim \pi R_{o}^2 \approx
10^{-25}\cm^2.$ and $v = \sqrt{2T/m_p}$ is baryon thermal
velocity. Then ANO-helium gas decouples from plasma and radiation
and plays the role of dark matter, which starts to dominate in the
Universe at $T_{RM}=1 \eV$.

The composite nature of ANO-helium makes it more close to warm
dark matter. The total mass of $(OHe)$ within the cosmological
horizon in the period of decoupling is independent of $S_5$ and
given by $$M_{od} = \frac{T_{RM}}{T_{od}} m_{Pl}
(\frac{m_{Pl}}{T_{od}})^2 \approx 2 \cdot 10^{42} \g = 10^9
M_{\odot}. $$ O-helium is formed only at $T_{o} = 60 \keV$ and the
total mass of $OHe$ within cosmological horizon in the period of
its creation is $M_{o}=M_{od}(T_{o}/T_{od})^3 = 10^{37} \g$.
Though after decoupling Jeans mass in $(OHe)$ gas falls down $M_J
\sim 3 \cdot 10^{-14}M_{od}$ one should expect strong suppression
of fluctuations on scales $M<M_o$ as well as adiabatic damping of
sound waves in RD plasma for scales $M_o<M<M_{od}$. It provides
suppression of small scale structure in the considered model. This
dark matter plays dominant role in formation of large scale
structure at $k>1/2$.

The first evident consequence of the proposed scenario is the
inevitable presence of ANO-helium in terrestrial matter, which is
opaque for $(ANOHe)$ and stores all its in-falling flux. If its
interaction with matter is dominantly quasi-elastic, this flux
sinks down the center of Earth. If ANO-helium regeneration is not
effective and $\Delta$ remains bound with heavy nucleus $Z$,
anomalous isotope of $Z-2$ element appears. This is the serious
problem for the considered model.

Even at $k=1$ ANO-helium gives rise to less than 0.1 \cite{I,lom}
of expected background events in XQC experiment \cite{XQC}, thus
avoiding for all $k \le 1$ severe constraints on Strongly
Interacting Massive particles SIMPs obtained in
\cite{McGuire:2001qj} from the results of this experiment. In
underground detectors $(ANOHe)$ ``molecules'' are slowed down to
thermal energies far below the threshold for direct dark matter
detection. However, $(ANOHe)$ destruction can result in observable
effects. Therefore a special strategy in search for this form of
dark matter is needed. An interesting possibility offers
development of superfluid $^3He$ detector
\cite{Winkelmann:2005un}. Due to high sensitivity to energy
release above ($E_{th} = 1 \keV$), operation of its actual few
gram prototype can put severe constraints on a wide range of $k$
and $S_5$.

At $10^{-3}<k<0.02$ $U$-baryon abundance is strongly suppressed
\cite{lom,KPS}, while the modest suppression of primordial $N$
abundance does not exclude explanation of DAMA, HEAT and EGRET data
in the framework of hypothesis of 4th neutrinos but makes the effect
of $N$ annihilation in Earth consistent with the experimental data.

%%%%%%%%%%%%%%%%%%%%%%%%%%%%%%%%%%%%%%%%%%%%%%%%%%%%%%%%%%%%%%%%%%%%%%%%%%%%%%%%%%%
%%%%%%%%%%%%%%%%%%%%%%%%%%%%%%%%%%%%%%%%%%%%%%%%%%%%%%%%%%%%%%%%%%%%%%%%%%%%%%%%%%%
\section{Signatures for $U$-hadrons in accelerator experiments}

Metastable $U$-quark within a wide range of expected mass can be searched on LHC and Tavatron.
In spite on that its mass can be quite close to that of $t$-quark, strategy of their search
should be completely different. $U$-quark in framework of considered model is metastable and will form
metastable hadrons at accelerator contrary to $t$-quark.

\begin{figure}
\includegraphics[scale=0.6]{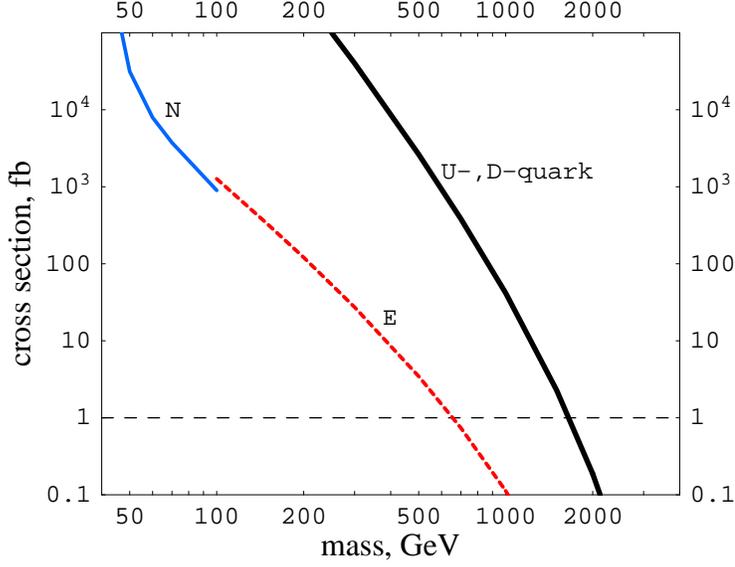}
\caption{Cross sections of production of 4th generation particles (N, E, U, D) at LHC.
Horizontal dashed line shows approximate level of sensitivity to be reached after first year of
LHC operation.}
\label{crossec}
\end{figure}

Detailed analysis of possibility of U-quark search requires quite deep understanding
of physics of interaction between metastable U-hadrons and nucleons of matter.
However, strategy of U-quark search can be described in general outline,
by knowing mass spectrum of U-hadrons, (differential) cross sections of their production.
LHC certainly will provide a better possibility for U-quark search than Tevatron.
Cross section of U-quark production in pp-collisions at the center mass energy 14 TeV
is presented on the Fig. \ref{crossec}. For comparison, cross sections of 4th generation
leptons are shown too. Cross sections of U- and D- quarks does not virtually differ.

Heavy metastable quarks will be produced with high transverse
momentum $p_T$, velocity less than speed of light. In general,
simultaneous measurement of velocity and momentum enables
to find the mass of particle. Information on ionization
losses is, as a rule, not so good thereto. All these features are
typical for any heavy particle, while there can be subtle
differences in the shapes of its angle-, $p_T$-distributions,
defined by concrete model which it predicts.

It is peculiarities of
long-lived hadronic nature what can be of special importance for
clean selection of events of U-quarks creation. U-quark can form a
whole class of U-hadronic states which can be perceived as stable
in condition of experiment contrary to their relics in Universe.
However, as we pointed out, double, triple U-hadronic states cannot
be virtually created in collider. Many other hadronic states whose
lifetime is $\gsim 10^{-7}$ s should look like stable. In the
Table 2 expected mass spectrum of U-hadrons, obtained with the help of
code Pythia \cite{pythia}, is presented.
\begin{table}
\caption{Mass spectrum and relative yields in LHC for U-hadrons. The same is for charged conjugated states.}
\begin{tabular}{|p{1.3in}|p{1.3in}|p{1.2in}|p{0.55in}|} \hline
%\newline
& {\small Difference between the masses of U-hadron and U-quark, GeV} &
\multicolumn{2}{|p{1.9in}|}{\small Expected yields (in the right columns the yields of long-lived states are given)} \\ \hline
$\left\{U\bar{u}\right\}^{0} ,\left\{U\bar{d}\right\}^{+} $ & 0.330  & \multicolumn{2}{|p{1.8in}|}{39.5(3)\%, 39.7(3)\%} \\ \hline
$\left\{U\bar{s}\right\}^{+} $ & 0.500  & \multicolumn{2}{|p{1.8in}|}{11.6(2)\%} \\ \hline
$\left\{Uud\right\}^{+} $ & 0.579  & 5.3(1)\% & %\newline
7.7(1)\% \\ \hline
$\left\{Uuu\right\}_{1}^{++} ,\left\{Uud\right\}_{1}^{+} ,$ $\left\{Udd\right\}_{1}^{0} $ & 0.771 & 0.76(4)\%, 0.86(5)\%, 0.79(4)\% &  \\ \hline
$\left\{Usu\right\}^{+} ,\left\{Usd\right\}^{0} $ & 0.805 & 0.65(4)\%, 0.65(4)\% & %\newline
1.51(6)\% \\ \hline
$\left\{Usu\right\}_{1}^{+} ,\left\{Usd\right\}_{1}^{0} $ & 0.930 & 0.09(2)\%, 0.12(2)\% &  \\ \hline
$\left\{Uss\right\}_{(1)}^{0} $ & 1.098 & \multicolumn{2}{|p{1.8in}|}{0.005(4)\%} \\ \hline
\end{tabular}
\end{table}

The lower indexes in notation of U-hadrons in the Table 2 mean (iso)spin ($I$) of the light quark pair.
From comparison of masses of different U-hadrons
it follows that all $I=1$ U-hadrons decay quickly emitting $\pi$-meson or $\gamma$-quantum,
except $(Uss)$-state. In the right column the expected relative yields are present.
Unstable $I=1$ U-hadrons decay onto
respective $I=0$ states, increasing their yields.

Firstly one makes a few notes. There are two
mesonic states being quasi-degenerated in mass: $(U\bar u)$ and $(U\bar d)$
(we skip here discussion of strange U-hadrons).
In interaction with medium composed of $u$ and
$d$ quarks transformations of U-hadrons into those ones containing
$u$ and $d$ are preferable (as it is the case in early Universe).
From these it follows, that created pair of $U\,\bar U$ quarks will fly
out from the vertex of pp-collision in form of U-hadron with
positive charge in about 60\% of such events
and with neutral charge in 40\% and in form of
anti-U-hadron with negative charge in 60\% and neutral in 40\%.
After traveling through detectors a few nuclear lengths from
vertex, U-hadron will transform in (roughly) 100\% to positively
charged hadron $(Uud)$ whereas anti-U-hadron will transform in 50\% to
negatively charged U-hadron $(\bar U d)$ and in 50\% to neutral
U-hadron $(\bar U u)$.

This feature will enable to discriminate the considered model of U-quarks from variety of alternative models,
predicting new heavy stable particles.

 Note that if the mass
of Higgs boson exceeds $2m$, its decay channel into the pair of stable
$Q \bar Q$ will dominate over the $t \bar t$, $2W$, $2Z$ and
invisible channel to neutrino pair of 4th generation
\cite{nuHiggs}. It may be important for the strategy of heavy
Higgs searches.

%%%%%%%%%%%%%%%%%%%%%%%%%%%%%%%%%%%%%%%%%%%%%%%%%%%%%%%%%%%%%%%%%%%%%%%%%%%%%%%%%%%%%
%%%%%%%%%%%%%%%%%%%%%%%%%%%%%%%%%%%%%%%%%%%%%%%%%%%%%%%%%%%%%%%%%%%%%%%%%%%%%%%%%%%%%
\section{\label{discussion} Conclusion}

To conclude, the existence of hidden stable or metastable quark of 4th
generation
can be compatible with the severe experimental constraints on
the abundance of anomalous isotopes in Earths atmosphere and ground
and in cosmic rays, even if the lifetime of such quark exceeds
the age of the Universe. Though the primordial abundance
$f= r_4/r_b$ of hadrons, containing such quark (and antiquark)
can be hardly less than $f \sim 10^{-10}$ in case of charge symmetry, their
primordial content
can strongly decrease in dense astrophysical objects (in the Earth, in
particular)
owing to the process of recombination, in which hadron, containing quark,
and hadron, containing antiquark, produce unstable charmonium-like
quark-antiquark state.

To make such decrease effective, the equal number density of
quark- and antiquark-containing hadrons should be preserved.
It appeals to a dynamical mechanism, preventing segregation
of quark- and antiquark- containing hadrons. Such mechanism,
simultaneously providing strict charge symmetry of quarks
and antiquarks, naturally arises, if the 4th generation
possesses new strictly conserved U(1) gauge ($y$-) charge.
Coulomb-like $y$-charge long range force between quarks and
antiquarks naturally preserves equal number densities
for corresponding hadrons and dynamically supports the
condition of $y$-charge neutrality.

It was shown in the present paper that if $U$-quark is the lightest
quark of the 4th generation, and the lightest free $U$-hadrons are
doubly charged $(UUU)$- and $(UUu)$-baryons and electrically neutral
$(U \bar u)$-meson, the predicted abundance of anomalous helium in
Earths atmosphere and ground as well as in cosmic rays is below the
existing experimental constraints but can be within the reach for
the experimental search in future. To realize this possibility
nuclear binding of all the $(Uud)$-baryons with primordial helium is
needed, converting potentially dangerous form of anomalous hydrogen
into less dangerous anomalous lithium. Then the whole cosmic
astrophysics and present history of these relics are puzzling and
surprising, but nearly escaping all present bounds.

Searches for anomalous isotopes in cosmic rays and at accelerators
were performed during last years. Stable doubly charged $U$
baryons offer challenge for cosmic ray and accelerator
experimental search as well as for increase of sensitivity in
searches for anomalous helium. In particular, they seem to be of
evident interest for cosmic ray experiments, such as PAMELA and
AMS02. $+2$ charged $U$ baryons represent the low $Z/A$ anomalous
helium component of cosmic rays, whereas $-2$ charged $\bar U$
baryons look like anomalous antihelium nuclei. In the baryon
asymmetrical Universe the predicted amount of primordial single
$\anti Uud$ baryons is exponentially small, whereas their secondary
fluxes originated from cosmic ray interaction with the galactic
matter are predicted at the level, few order of magnitude below
the expected sensitivity of future cosmic ray experiments. The
same is true for cosmic ray $+2$ charged $U$ baryons, if $U$-quark
lifetime is less than the age of the Universe and primordial $U$
baryons do not survive to the present time.

The models of quark interactions favor isoscalar
$(Uud)$ baryon to be the lightest among the 4th generation baryons
(provided that $U$ quark is lighter, than $D$ quark, what also may
not be the case). If the lightest $U$-hadrons have electric charge
$+1$ and survive to the present time, their abundance in Earth
would exceed the experimental constraint on anomalous hydrogen.
This may be rather general case for the lightest hadrons of the
4th generation. To avoid this problem of anomalous hydrogen
overproduction the lightest quark of the 4th generation should
have the lifetime, less than the age of the Universe. Another
possible solution of this problem, using double and triple $\bar
U$ baryons $\anti UUU$ and
catalysis of $(Uud)$ annihilation in atom-like bound systems
$^4He\anti UUU$ is considered in \cite{BFK}.

However short-living are these quarks on the cosmological
timescale in a very wide range of lifetimes they should behave as
stable in accelerator experiments. For example, with an
intermediate scale of about 10$^{11} \GeV$  (as in supersymmetry
models \cite{Benakli}) the expected lifetime of $U$-(or $D$-)
quark $\sim 10^6$ years is much less than the age of the Universe
but such quark is practically stable in any collider experiments.

First year operation of the accelerator LHC has good discovery potential for
U(D)-quarks with mass up to 1.5 TeV. U-hadrons born at accelerator will
distinguish oneself by high $p_t$, low velocity, by effect of a charge flipping
during their propagation through the detectors. All these features
enable strongly to increase efficiency of event selection from not only
background but also from alternative hypothesis.

In the present work we studied effects of 4th generation
having restricted our analysis by the processes
with 4th generation quarks and antiquarks. However,
as we have mentioned in the Introduction in the considered
approach absolutely stable neutrino of 4th generation
with mass about 50 GeV also bears $y$-charge.
The selfconsistent treatment of the cosmological
evolution and astrophysical effects of $y$-charge
plasma of neutrinos, antineutrinos, quarks and antiquarks of
4th generation in charge symmetric case will be the subject of special studies.
An attempt of such a treatment has been undertaken in the case of charge
asymmetry, described in this paper.

We believe that a tiny trace of heavy hadrons as anomalous helium
and stable neutral $\OHe$ and mesons\footnote{Storing these charged
and neutral heavy hadrons in the matter might influence its $e/m$
properties, leading to the appearance of apparent fractional charge
effect in solid matter \cite{BFK}. The present sensitivity for such
effect in metals ranges from $10^{-22}$ to $10^{-20}$ .} may be
hidden at a low level in our Universe ($\frac{n_U}{n_b} \sim
10^{-10} - 10^{-9}$) and even at much lower level here in our
terrestrial matter a density $\frac{n_U}{n_b} \sim 10^{-23}$ in case
of charge symmetry. There are good reasons to bound the 4th quark
mass below TeV energy. Therefore the mass window and relic density
is quite narrow and well defined, open to a final test.

In case of charge asymmetry of 4th generation quarks, a nontrivial
solution of the problem of dark matter (DM) can be provided due to
neutral $\OHe$-like U-hadrons states (ANO-helium in case of
$y$-interaction existence). Such candidates to DM have many
interesting implications in BBN, large scale structure of Universe
and physics of DM \cite{I,lom,KPS,FKS,Majorana,KK,Bled07,DM08}. It
should catalyze new types of nuclear transformations, reminding
alchemists' dream on the philosopher's stone. It challenges direct
search for species of such composite dark matter and its
constituents. A very low probability for their existence is strongly
compensated by the expectation value of their discovery.

\section*{Acknowledgements}

We are grateful to G. Dvali for reading the manuscript and important recommendations

\end{document}